\documentclass[10pt,onecolumn]{elsart}
\usepackage[a4paper]{anysize}
\usepackage{rotating}
\usepackage{epsfig}
\usepackage{graphicx}
\usepackage{caption}
\usepackage{amsmath}
\usepackage{latexsym}
\usepackage{lscape}
\begin{document}

\begin{frontmatter}

\title{Studies of aging and HV break down problems during development and operation of MSGC 
and GEM detectors for the Inner Tracking System of \mbox{HERA-B} \thanksref{bmbf} }
\thanks[bmbf]{Work supported by 
Bundesministerium f\"ur Bildung und Forschung, Deutschland, and the Swiss National Science Foundation.}
\author[desy]{Y. Bagaturia}
\author[si,bonn]{O. Baruth}
\author[hd,sap]{H.B. Dreis}
\author[hd]{F. Eisele}
\author[si]{I. Gorbunov}
\author[hd]{S. Gradl}
\author[hd]{W. Gradl}
\author[hd,hp]{S. Hausmann}
\author[hd,zunew]{M. Hildebrandt}
\author[hd,cern]{T. Hott}
\author[si,bosch]{S. Keller}
\author[hd]{C. Krauss}
\author[desy,lip]{B. Lomonosov}
\author[desy,lip]{M. Negodaev}
\author[hd,mc]{C. Richter}
\author[zu]{P. Robmann}
\author[hd,desynew]{B. Schmidt}
\author[hd,zunew]{U. Straumann}
\author[zu]{P. Tru\"ol}
\author[hd,siemens]{S. Visbeck}
\author[zu]{T. Walter}
\author[hd]{C. Werner}
\author[si]{U. Werthenbach}
\author[si]{G. Zech}
\author[si]{T. Zeuner}
\author[hd,zunew]{M. Ziegler}
\address[hd]{Universit\"at Heidelberg, Germany}
\address[si]{Universit\"at Siegen, Germany}
\address[zu]{Universit\"at Z\"urich, Switzerland}
\address[desy]{DESY, Hamburg}
\address[zunew]{now at Universit\"at Z\"urich}
\address[cern]{now at CERN, Geneva}
\address[desynew]{now at DESY, Hamburg}
\address[bonn]{now at Universit\"at Bonn}
\address[sap]{now at SAP, Walldorf}
\address[hp]{now at Hewlett-Packard, Hamburg}
\address[bosch]{now at Bosch, Stuttgart}
\address[siemens]{now at Siemens, M\"unchen}
\address[mc]{now at McKinsey, Frankfurt}
\address[lip]{On leave of absence from LIP Moscow, Russia}


\begin{abstract}
  The results of five years of development of the inner tracking
  system of the HERA-B experiment and first experience from the data taking period of the year 2000
  are reported. The system contains 184 chambers,
  covering a sensitive area of about (20 $\times$ 20) cm$^{2}$ each. The detector is based on
  microstrip gas counters (MSGCs) with diamond like coated (DLC) glass wafers and gas electron multipliers (GEMs).
  The main problems in the development phase were gas discharges in intense hadron
  beams and aging in a high radiation dose environment. The observation of gas discharges
  which damage the electrode structure of the MSGC
  led to the addition of the GEM as a first amplification step.
  Spurious sparking at the GEM cannot be avoided completely. 
  It does not affect the GEM itself but can
  produce secondary damage of the MSGC if the electric field between the GEM
  and the MSGC is above a threshold depending on operation conditions.
  We observed that aging does not only depend on the dose but also on the
  spot size of the irradiated area. Ar-DME mixtures had to be abandoned
  whereas a mixture of 70\% Ar and 30\% CO$_2$ showed no serious aging effects up to about
  40 mC/cm deposited charge on the anodes. X-ray measurements
  indicate that the DLC of the MSGC is
  deteriorated by the gas amplification process. As a consequence,
  long term gain variations are expected. The Inner Tracker has
  successfully participated in the data taking at HERA-B during summer
  2000.
\end{abstract}

\end{frontmatter}

\section{Introduction}

The \mbox{HERA-B} experiment \cite{herab} was designed with the goal to measure CP
violation in the B-system. It started operation at DESY, Hamburg in
spring 2000. Neutral B-Mesons are produced by interactions of 920 GeV
protons on a stationary nuclear target followed by a magnetic
spectrometer. Since the B cross section is very low, the experiment was planed for an interaction rate
of 40 MHz to obtain an acceptable B production rate. 
The detectors therefore were designed to withstand the corresponding high particle rates and
high radiation levels.

To cope with the high particle flux which drops roughly as one over the
distance from the beam axis squared, the main tracking system has been
subdivided into two parts with different rate capabilities: the Inner
Tracker (ITR) near the beam pipe and at larger distance the Outer
Tracker (OTR) consisting of drift chambers composed of honeycomb structures.

\begin{figure}[tbp]
\begin{center}
\epsfig{clip=,file=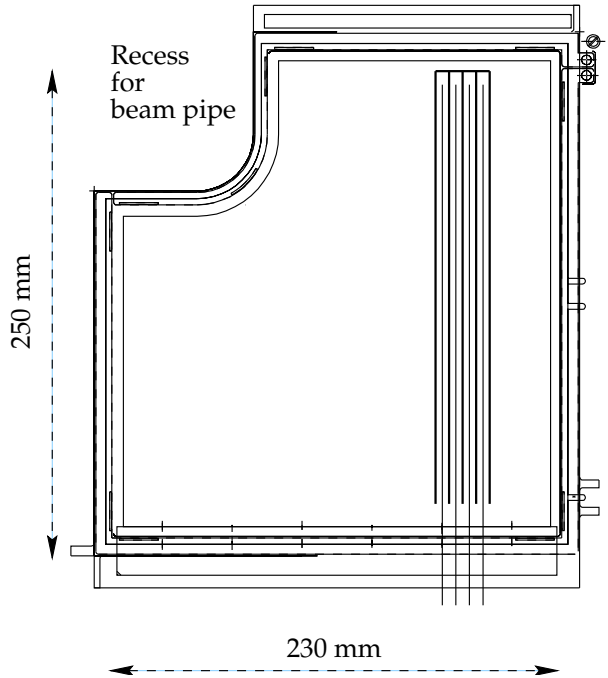,width=8cm}
\end{center}
\caption{Sketch of a MSGC chamber indicating size and shape of the
  wafer.}
\label{detlay}
\end{figure}

The outer dimensions of the sensitive area of the ITR are given by the
requirement that the occupancy per channel of the OTR does not exceed
a critical value of 20\%. The high intensity area can be
covered by four MSGC detectors with wafer
size of \mbox{25$\times$25 cm$^{2}$} which can be realized in industry.
The basic geometry of a MSGC detector is shown
in Figure \ref{detlay}. The detectors have a round recess which permits to position four
MSGCs around the beam pipe.
They have to sustain a particle flux of up to \mbox{2$\cdot$10$^{6}$/cm$^{2}$/s}
and radiation doses of up to 1 Mrad/year. These requirements
are very similar to those of the inner tracking in the ATLAS and CMS
experiments for LHC and are a big challenge for the tracking
detector technology.

The total \mbox{HERA-B} ITR system requires 46 detector planes, each
consisting of four chambers.  Each of the 184 chambers has 752 anode
strips leading to a total of about 140000 readout channels.  This
basic design has remained unchanged, the detector technology, however,
had to be modified several times as a consequence of R\&D work and,
especially, the results of beam tests.

At the time of approval of the experiment in 1995, the MSGC technology was
claimed to be ready for high rate applications \cite{rd28,ard28,CMSLOI}. Very
promising test results had been obtained and published with the claim that
MSGCs on ordinary borosilicate glass wafers can be operated at high rates
with good stability and homogeneous response. Moreover, aging tests with
Ar-DME (argon-dimethylether) mixtures and clean gas systems were reported which were interpreted
to demonstrate that these detectors can survive up to \mbox{10 years} of LHC
operation (equivalent to about the same time at \mbox{HERA-B}) without severe
problems. Finally, the production of large size wafers with small error rate
at rather low cost was promised.

The construction of an inner tracker for \mbox{HERA-B} based on the MSGC technology
therefore looked rather straight forward when the experiment was approved.
In the course of our development work we faced however several serious and
unexpected problems:

\begin{itemize}
\item  The production of large size MSGC wafers \mbox{25$\times$25 cm$^{2}$}
with acceptable quality and a small fraction of broken anodes and cathodes
turned out to be difficult. After extensive R\&D at the company IMT\footnote{%
IMT, Greifensee, Switzerland}, series production of these wafers now has
high yield with less than 5\% interrupted anodes and no shorts.

\item  MSGCs based on bare glass (DESAG 263) wafers as proposed were unable
to satisfy the requirements on gain homogeneity, rate stability and radiation tolerance.
 These problems were solved by the introduction of ``diamond
 like coating'' (DLC) which provides a well defined electronic surface conductivity of
the substrate.

\item MSGCs with DLC exhibit gas discharges
  between anodes and cathodes when operated in a hadron beam at the
  required high gas gain. Every discharge damages electrodes
  and in the long run this effect leads to an intolerable number of
  broken anodes. This problem could be solved by the introduction of
  an additional amplification step using the GEM technology
  \cite{sauligem}.

\item The MSGC/GEM technology is vulnerable. We have experienced a
series of problems with gas discharges at the GEM which subsequently 
lead to damages of the MSGC electrodes.

\item Moreover, we were repeatedly faced with aging problems due to traces of
impurities which are very hard to avoid reliably during production of a
large series and during long term operation. Above all, the Ar-DME mixture
originally foreseen as counting gas in most MSGC developments showed
rapid aging in large area X-ray and test beam
illumination of detectors. Finally, an Ar-CO$_{2}$ mixture had to be used.
\end{itemize}

In the following, major development steps and the most important results are
described.

\section{The MSGC detector}

\subsection{Detector geometry}

The electrode structure chosen for the MSGC (type A) has a pitch of \mbox{300 $\mu$m}, anode
width of \mbox{10 $\mu$m} and an anode cathode gap of \mbox{60 $\mu$m}. 48 chambers (type B)
located at the downstream end of the tracking have to cover a larger
sensitive area. They have a pitch of \mbox{350 $\mu$m} allowing to use the same
readout configuration as for the smaller chambers. The pitch is a compromise
between cost of the electronics and maximum tolerable occupancy. The
relatively large anode width was chosen to minimize the number of anodes
interrupted during production. The maximum length of an anode in the
experiment is \mbox{23 cm}. The height of the drift gap was chosen to be 3.3 mm. This
guarantees that all primary electrons have reached the anodes within 60 ns
even for slow Ar-DME mixtures and that the anode signal occupies only one
bunch crossing. Subsequent bunch crossings are 96 ns apart.

The high voltage is distributed via a resistor network
to groups of 16 cathodes. In this way, a short between an anode and a
cathode strip leads to an inefficient region of about 2 \% of the total
chamber area. 

\subsection{Wafers and coating}

Experience with first prototypes of MSGCs \cite{visb96} showed that bare glass with 
its relatively low ionic conductivity cannot be used in a high rate and high 
dose environment.  Local gain variations up to a factor of three were found 
even at low counting rates. The gas gain was rate dependent and 
the detectors did not survive an X-ray dose equivalent to about half a
\mbox{HERA-B} year. Glass with
electronic conductivity was excluded both by cost and its short radiation
length. The obvious solution was to coat the surface with an electronically
conducting compound. Many different coatings have been tested by various
groups. Finally, DLC proved to be adequate.

DLC surfaces had been developed simultaneously by the CERN group
together with the company SURMET\footnote{%
SURMET Corporation, Burlington, MA, USA.}, and by the Siegen group together
with a Fraunhofer Institut\footnote{%
Fraunhofer Institut f\"{u}r Schicht- und Oberfl\"{a}chentechnik,
Braunschweig, Germany.}. The two coatings are very similar in their chemical
composition. They consist of mixtures of the elements C, Si, H, N deposited
on the glass by chemical vapor deposition (CVD) using gas mixtures of CH$
_{4} $, SiH$_{4}$, N$_{2}$. In fact, the coating consists mainly of
amorphous carbon and has very little diamond content, if any. The relative
abundance of nitrogen and hydrogen and the production temperature determine
the electric conductivity. Silicon improves the adhesion of the coating
layer which has a typical thickness of 80 nm. The resistivity of the coating
can be increased by tempering the wafers. Tempering also increases the
stability of the resistivity with time. Wafers coated by SURMET were
delivered with typical resistivity of $10^{15}$ $\Omega / \Box$ whereas the
resistivity of the Fraunhofer coating was originally lower by one to two orders
of magnitude.

The SURMET coating had reasonable electrical properties and very good
chemical stability. The mechanical quality of the large area coating was
however not satisfactory most likely due to problems in handling and
cleaning. For the main series production the DLC was therefore
produced by the Fraunhofer Institut on AF45 glass (alkali ``free'', produced
by DESAG ). This type of glass with little alkali content was chosen to
reduce the risk of electrolytic modifications of the substrate, caused by
the continuous alkali ion current over several years \cite{lyon95}. The coating has an
average surface conductivity of $10^{14}$ $\Omega / \Box$ which varies over
the area by about 20 \%.

The lithographic production of the electrode pattern was realized by the
company IMT on the coated wafers of 0.4 mm thickness and outer dimensions of
about \mbox{30$\times$30 cm$^{2}$}.

To avoid overlapping hits in subsequent bunch crossings, we have to avoid electronic
broadening of the signals propagating along the anode beyond about 50 ns.
The signal shape at the amplifier input depends on the distance of the avalanche 
to the amplifier end of the anode and on the product of anode-cathode resistivity and capacitance. 
Delta shaped voltage pulses generated at the far end of the anode 
produce exponentially decaying signals at the amplifier input with a time 
constant $\tau=4RC/\pi^{2}$, with $R$ the total anode resistance and $C$ the anode cathode
capacitance. Thus the signal width is proportional to the strip length squared. 
For the given anode strip length of 23 cm and anode cathode capacitance of 20 pF 
an anode material of 
resistivity\footnote{The actual resistivity of lithographically produced electrodes is
 usually much higher than the nominal table values for bulk material.} below \mbox{250
$\Omega$/cm} is required.
This requirement excludes nickel and chromium and points to aluminium
or gold. Aluminium is preferable for lithography and has also rather
good resistance to sparks (see Section 2.4).  Aging tests with
aluminium wafers however were discouraging (see Section 4).  The
aluminium option was therefore abandoned. A \mbox{500 nm} thick gold
MSGC pattern was produced by a lift-off technique.

The thickness of the different wafer materials is summarized in Table 1.

\begin{table}\centering
\begin{tabular}{|l|c|c|c|c|}

\hline
material   & AF45             & CVD            & titanium           & gold                \\
           & glass substrate  &     coating    & adhesion layer     & electrode structure \\
\hline
thickness  & 400 $\mu m$      & 0.080 $\mu m$  & 0.050 $\mu m$      & 0.500 $\mu m$       \\
\hline

\end{tabular}

\caption{\it Summary of the material thickness for the different wafer layers. \label{wafermat}}
\end{table}

After major improvements of the spinning techniques, of cleaning and
etching processes and semi-automatic handling of the wafers, IMT was able to
produce with high yield wafers with less than 5 \% broken anodes.
The number of faults was measured by a specially designed electrical
test station \cite{unizuerich}. If a short was found, the wafer was repaired
by etching away the short. Accepted wafers had no shorts and an average of 2 \%
interrupted anodes.

\subsection{Operation characteristics}

Small size prototypes \mbox{10 $\times$ 10 cm$^2$} of MSGCs with coated wafers were systematically tested.
The homogeneity
of the gas gain over the full area showed variations of less than 20 \%. 
During a long series of aging tests, chambers operated 
with an argon dimethylether mixture of 50 \% Ar and \mbox{50 \%} DME (Ar-DME 50/50)
were illuminated with X-rays over typical areas of up to \mbox{113 mm$^{2}$}.   
The chambers showed constant gain up to charge depositions of \mbox{80
mC/cm} corresponding to more than 10 years of \mbox{HERA-B} operation. Rate tests demonstrated
that the detectors could be operated at gas gains of 3500 up to rates of \mbox{$%
10^{5}$} absorbed X-quanta /mm$^{2}$/s with a loss of pulse
height of less than 40\%. In laboratory
measurements with a $\beta$ source, efficiencies above 99\% were achieved.
This type of MSGCs therefore promised good performance for \mbox{HERA-B}.

In the year 1996 a beam test at PSI\footnote{Paul Scherrer Institute, Villingen, Switzerland.}
with intense hadron beams was performed to measure absolute efficiencies for MIPs and
the spatial resolution in a realistic environment. The test used small size 
\mbox{10 $\times$ 10 cm$^{2}$} detectors and a pion beam with momentum of 150 MeV/c 
which allowed
particle fluxes in the beam spot up to 3000 particles/mm$^{2}$/s.

\begin{figure}[tbp]
\begin{center}
\includegraphics[width=10cm]{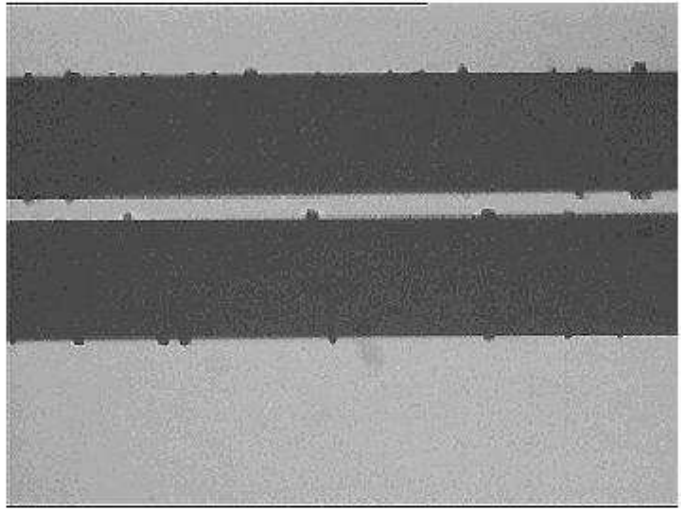}
\end{center}
\caption{Photograph of MSGC electrodes after operation of the chamber in a
pion beam. Characteristic defects are observed at the electrode borders.}
\label{mousebite}
\end{figure}

Already the very first day of operation we observed a severe problem which
was not encountered in the laboratory. The cathode current showed large
spikes, at the same time anodes got lost and sometimes pre-amplifier
channels were destroyed. Inspection of the
electrode structure in the laboratory under the microscope showed
characteristic defects. In the area of the beam spot marks are found on the
gold electrodes where the gold is evaporated both from the anode and from the
cathode just opposite to each other. A typical photograph is shown in Figure 
\ref{mousebite}. Obviously gas discharges had occurred which sometimes were
so intense that they cut an anode strip completely. 
The lifetime of the chambers in the HERA-B environment with the foreseen 
technology would have been only months.

Since discharges had never been observed in intense X-ray beams at exposures
with much higher charge deposition rate, it was suspected that the origin
of the discharges were strongly ionizing particles like nuclear fragments
produced in hadronic beams through interaction with detector material.

A series of laboratory tests where large local ionization was created by $%
\alpha $ particles was carried out to study this effect.

\subsection{Gas discharges by strongly ionizing
  particles}

The observed sparking phenomena in MSGC and MSGC/GEM detectors have
partially been published \cite{sparksschmidt,nimkeller}. Here a
summary of the results is given. Systematic measurements are difficult because
the sparking process damages the electrodes and thus modifies the operation conditions
during a measurement. Repeating measurements with the same detector often is impossible.

In order to get ionization by heavy particles the counting gas was guided
through a stainless steel volume filled with thorium powder. Thorium decays
into radium which is transported with the gas into the MSGC gas volume where
it emits $\alpha$ particles. A typical $\alpha$ particle counting rate of 10
Hz could be achieved this way. The MSGC detectors were operated at nominal
voltages and could also be illuminated in addition with an intense X-ray
beam. Gas discharges were detected by detection of large induced signals on the
cathodes. Alternatively, direct irradiation with a collimated $\alpha$%
-source through a thin window was used to measure local effects.

\begin{figure}[tbp]
\centerline{\includegraphics[width=12cm]{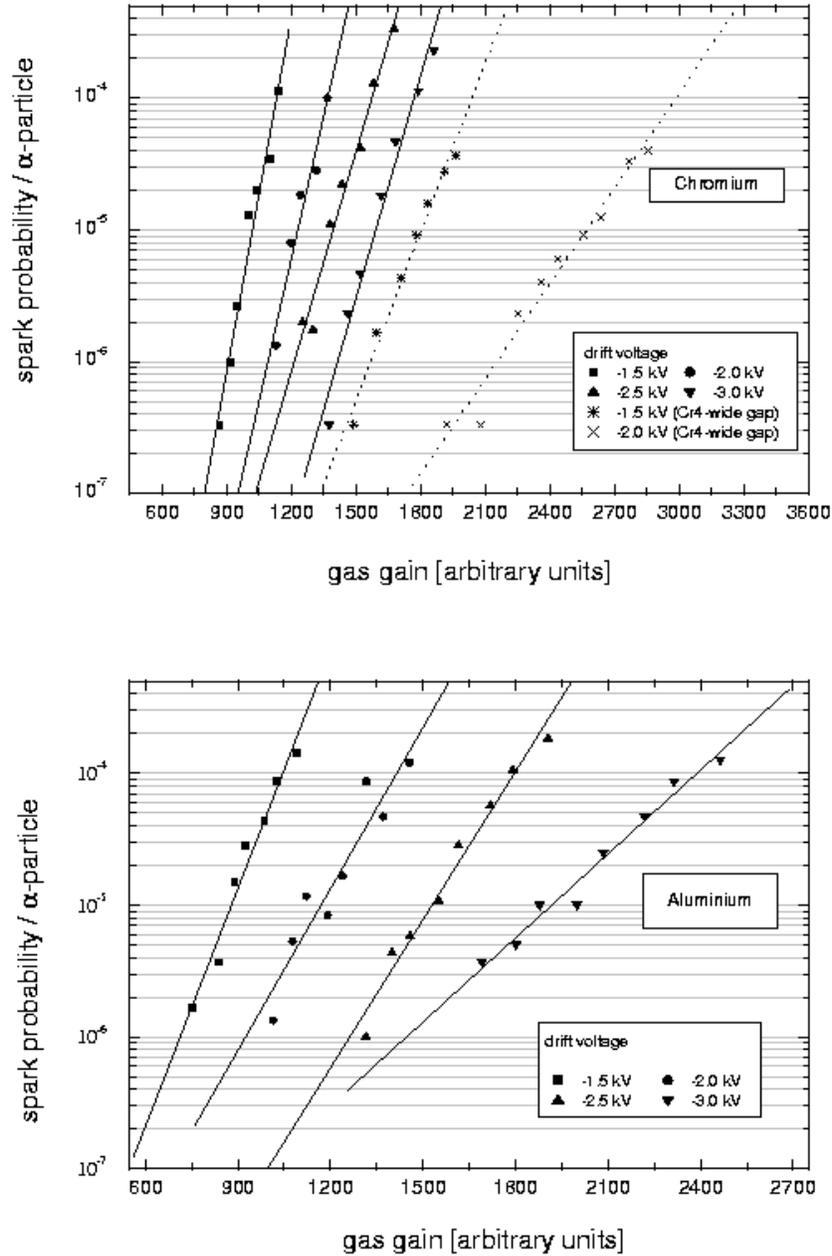}}
\caption{Spark probability per $\alpha$ particle as a function of the cathode and 
drift voltages measured with a \mbox{0.15 $\mu$m} chromium and a \mbox{1 $\mu$m} aluminium
    structure.}
\label{alphasparks}
\end{figure}

A characteristic result from the second  method is presented in \mbox{Figure \ref
{alphasparks}} which shows the spark rate as a function of the cathode and
drift voltages measured with a \mbox{0.15 $\mu $m} thick chromium structure and with
\mbox{1 $\mu $m} thick aluminium electrodes. We
observe a strong exponential rise of the spark rate with the cathode voltage which is 
responsible for the field near the surface and a much weaker
dependence on the drift voltage. At fixed gas gain, high drift fields are
favorable.

The effect of the width of the anode cathode gap was studied by increasing
it by a factor two to \mbox{120 $\mu$m}. At fixed gain the spark rate is smaller (see Figure \ref
{alphasparks}),
but this positive effect is more than compensated by the more severe damage
per spark observed with the wide gap, where the energy stored in the
anode-cathode capacitance is much larger.

\begin{figure}[htbp]
\centerline{\includegraphics[width=12cm]{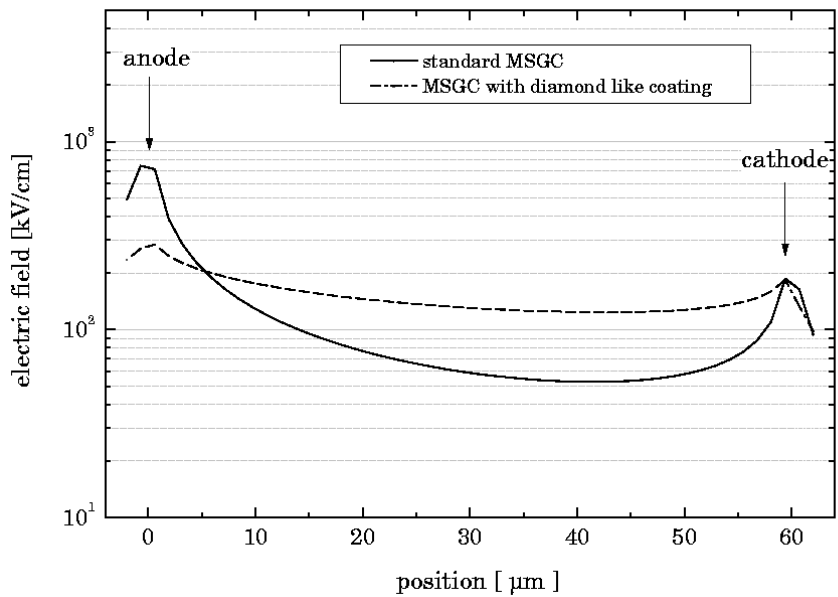}}
\caption{{Simulation of electric field for coated and uncoated MSGC.}}
\label{coatfield}
\end{figure}

The high sparking rate of the coated MSGCs is explained by the nearly constant
electric field component parallel to the surface near the surface, whereas
bare glass produces a strongly decreasing field near the anode, which
suppresses streamer discharges \cite{peskov,peskov1}. The electric fields between
the electrodes computed for a coated and an uncoated substrate are shown in
Figure \ref{coatfield}. Even though dynamical effects from the space charge 
created by the avalanche are neglected, the simulation illustrates  
the qualitative difference between the two configurations.

The explanation for the reduced sparking threshold for coated 
wafers was confirmed by experimental evidence that sparks were predominantly 
induced by ionization near the surface and is supported by the favorable effect 
of high drift fields at fixed gas gain.

The spark rate should mainly depend on the electric field configuration
and the geometry and be rather independent of the electrode material but we expected a
different robustness of various electrode metals against sparking. The hope
was that elements with high melting points and high binding energies like
rhodium or tungsten would be less vulnerable than gold. However, we did not observe any
substantial difference between Au, Cr\footnote{%
Chromium was excluded as a possible electrode material  because its
resistivity is too high. However, it is well suited for  systematic studies
of discharges in MSGCs.}, W and Rh and only a slightly better
performance of Al for a given resistivity of the anode \cite{kell98}. Aluminium electrodes
which can be produced with higher thickness than gold\ or rhodium sustained
a somewhat higher number of sparks before they break but this metal is
excluded because of its rather bad aging properties. Anodes with high
electric resistivity inhibit discharge of the full anode cathode capacitance
in a single spark. They show sequential discharges of low charge. The first spark 
discharges a few centimeters and produces a local potential drop. A subsequent spark occurs 
after the potential has again reached a critical value. This process repeats itself until 
the total charge stored in the anode cathode capacitance drops below a certain minimum. This
process limits the total energy release in the discharge and explains the fact that
chromium structures can survive more than 10$^{6}$ sparks without being
destroyed. The sequential discharges are well described by simulations of the electric circuitry \cite{nimkeller,kell98}.

Unfortunately, as explained in Section 2.2, anodes of high electric
resistance are incompatible with the requirement of short readout pulses.  
The geometry of the \mbox{HERA-B} detectors with long strip lengths and the correspondingly high
capacitance per channel of 25 pC (including the contribution from the
connections) leads to a white noise of the readout of 2250 
electrons per channel. Given our amplifier characteristics, a total gas gain 
exceeding 4000 is required to get good efficiency. With such a gas gain the discharge rate is unacceptably
high in intense hadron beams such that MSGC detectors of the \mbox{HERA-B} geometry
cannot be operated.

Ne-DME mixtures were reported to have higher primary ionization \cite{nedme}.
However, our tests have shown that the sparking probability at fixed gain
cannot be significantly improved within the uncertainties of the measurement of about 20 \%.

The sparking problem could only be solved by introducing a gas electron
multiplier (GEM)  foil \cite{sauligem} as a first amplification step and
thereby reducing the gain requirement for the MSGC.

\section{ MSGC-GEM detectors}

\subsection{Detector geometry}

\begin{figure}[tbp]
\begin{center}
\epsfig{clip=,file=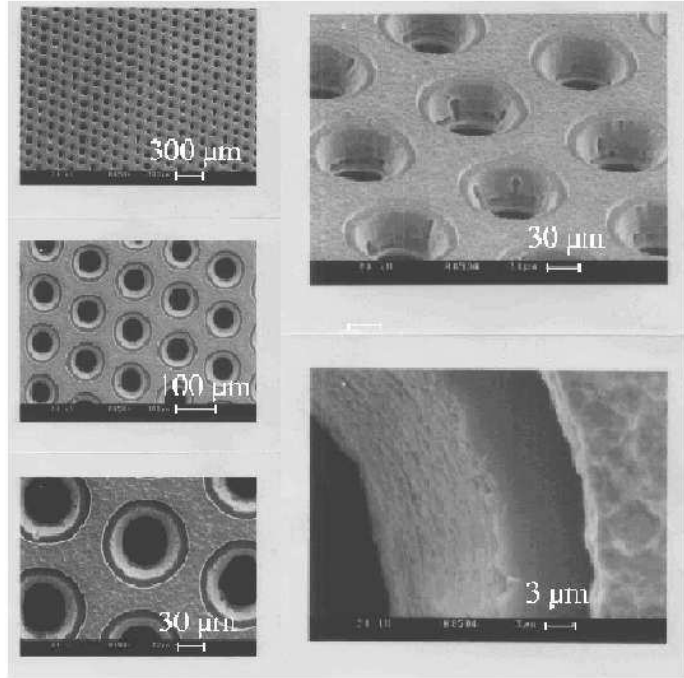,width=10cm}
\end{center}
\caption{{Photographs of a GEM with different enlargements}}
\label{topgem}
\end{figure}

\begin{figure}[tbp]
\begin{center}
\epsfig{clip=,file=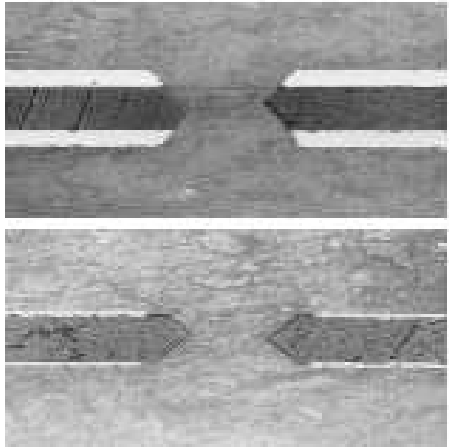,width=6cm}
\end{center}
\caption{{Cross section of GEM foil before and after final etching step.%
}}
\label{crossgem}
\end{figure}

The GEMs used for the ITR were produced at the CERN
workshop\footnote{CERN Surface Treatment Workshop, Geneva, Switzerland} using a
wet etching technique. They were made out of \mbox{50 $\mu $m} thick
polyimide (Kapton) foils coated with a \mbox{15 $\mu $m} copper layer on
each side. In a final
etching step the thickness of the copper layers was reduced to about \mbox{7 $\mu $%
m}. The conical holes have diameters of about \mbox{50 $\mu $m} in the
Kapton and \mbox{100 $\mu $m} in the copper. Photographs of the GEM and of its cross section
are shown in Figures \ref{topgem}, \ref{crossgem}. The holes are
arranged in a hexagonal lattice with a hole distance of \mbox{140 $\mu $m}
from center to center.

\begin{figure}[tbp]
\begin{center}
\epsfig{clip=,file=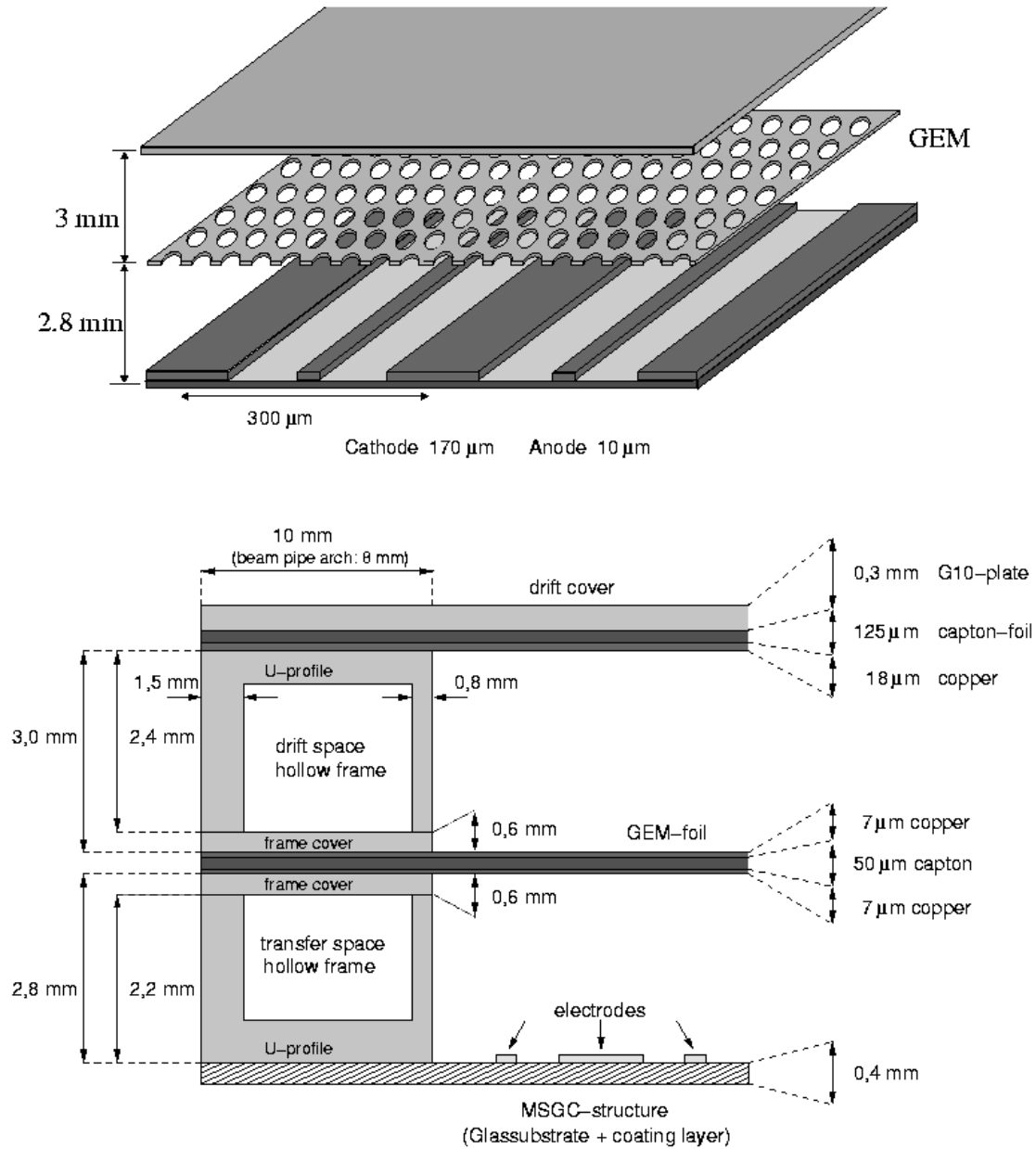,width=16cm}
\end{center}
\caption{Schematic view of a MSGC/GEM detector (top) and cross
section showing the frame (bottom).}
\label{msgcgemcross}
\end{figure}

The introduction of the GEM foil required a complete re-design of the
chambers. Figure \ref{msgcgemcross} illustrates the construction. The frames
which provide the mechanical stability to keep the GEM at constant distance
from the MSGC and distribute the gas, consist of two hollow epoxy pieces. The drift
gap above the GEM and the transfer gap below are 3.0 mm and 2.8 mm wide,
respectively.

The geometrical parameters of the detector are summarized in Table \ref{geom}.

\begin{table}\centering
\begin{tabular}{|l|c|}
\hline
\textbf{MSGC} &  \\ 
active area, type A & 528 cm$^{2}$ \\ 
active area, type B &  689 cm$^{2}$ \\ 
wafer thickness & 400 $\mu $m \\ 
anode width & 10 $\mu $m \\ 
anode cathode gap & 60 $\mu $m \\ 
pitch, type A & 300 $\mu $m \\ 
pitch type B & 350 $\mu $m \\ 
\hline
\textbf{GEM} &  \\ 
GEM thickness & 64 $\mu $m \\ 
thickness of GEM electrodes & 7 $\mu $m \\ 
hole diameter at center & 50 $\mu $m \\ 
hole diameter at electrodes & 100 $\mu $m \\ 
\hline
\textbf{Frame} &  \\ 
height of drift gap & 3 mm \\ 
height of transfer gap & 2.8 mm \\
\hline

\end{tabular}
\caption
{\it Summary of the geometrical detector parameters. \label{geom}}
\end{table}

\subsection{Operation characteristics}

Splitting of the gas amplification into two steps allowed to operate
the MSGC at moderate voltage. Typical voltage settings during the testing phase
were $U_{cath}=-520$ V, $\Delta U_{GEM}=420$ V, $U_{drift}=2.5$ kV which correspond 
to gas amplification factors of about 35 for the GEM and 250 for the MSGC. 
For comparison to reach the same amplification with a single MSGC structure,
one would have to apply about 100 V higher cathode voltages.
The detectors were tested intensively in the laboratory and at hadron
beams at PSI and in the HERA-B experiment.

The GEM shows the characteristic behavior of devices containing dielectric
materials. After switching on the HV, the gas amplification of the GEM rises
by a factor of about two within a few hours and stabilizes after a few days,
when the polyimide is fully polarized. The gain is rather uniform. It varies
by less than 20 \% over the full area of the chamber.
During irradiation the polyimide charges up and the gain increases by up
to 20 \%. The gain variations with time and rate can be inhibited by coating
the GEM foils \cite{gemcoat}. Since they are not critical for the operation
at HERA we avoided the complication, risk and cost of an additional production
step.

With the GEM alone, gas amplification up to \mbox{10$^{3}$} could be obtained
with photon irradiation before sparking started. 

A part of the ITR chambers are located inside the spectrometer magnet with a field
of 0.85 T. Initial worries that the GEM functionality would suffer in moderate
magnetic fields were not confirmed by a test at the nominal HERA-B field perpendicular
to the hole axis. No decrease of the gain was observed within 5 \% \cite{hild99}.

\subsection{GEM sparks inducing secondary sparks}

First laboratory tests at the PSI pion beam in 1998 of four MSGC/GEM
chambers  operated with Ar-DME 50/50 at a GEM gain of about 50 
gave satisfactory results \cite{dreis98}. A
second test  one year later, however, was similarly disastrous as our first MSGC
tests at a hadron beam. GEM sparks were observed which induced
secondary sparks at the microstrip structure, and sometimes led to
discharges between the GEM and the microstrip structure or between GEM
and the drift electrode. The electrodes of the MSGC were severely damaged.

\begin{figure}[tbp]
\begin{center}
\includegraphics[width=12cm]{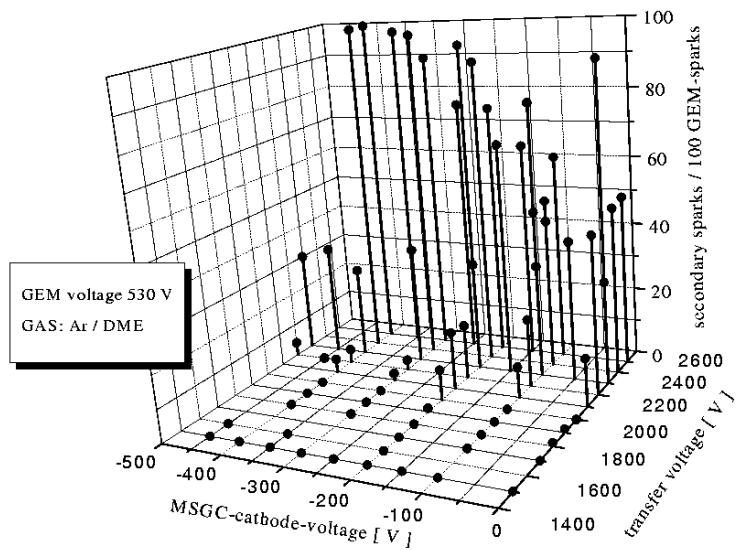}
\end{center}
\caption {Spark rate as a function of the cathode voltage and the voltage between GEM and the anodes. A transfer field of 2.5 kV/cm corresponds to }
\label{gemtransfer}
\end{figure}

In extensive laboratory studies the secondary sparking was
reproduced and investigated. The details of the complex sparking process
were never fully understood. However, it is obvious that a discharge of the
GEM with its high capacitance generates a huge amount of ions and electrons.
Even a small fraction of the electrons drifting towards the MSGC, with
additional gas amplification there, are likely to produce secondary sparks.
Thus the rate of those strongly depends on the strength of the transfer field. 
This was confirmed by
laboratory measurements presented in Figure \ref{gemtransfer} and it explained
also the different behavior of the chambers at the two PSI tests which were
performed at slightly different HV settings. 

A significant dependence of the secondary spark rate 
on the cathode voltage has not been observed (see Figure \ref{gemtransfer}). 

With the voltage between anodes and lower GEM plane limited to below 1250 V in the HERA-B experiment,
the MSGC/GEM detectors were able to cope with the intense hadron flux
without suffering from fatal damage by discharges (see Section 5).

\section{Choice of counting gas and aging studies}

Two gas mixtures were systematically studied, Ar-DME 50/50 and Ar-CO$_{2}$
70/30. The DME mixture has better quenching properties hence less
problems with gas discharges. It had also been extensively tested 
with respect to aging properties by several
groups, which consistently reported no aging
problems up to large charge depositions on the strips corresponding to many
years of LHC or \mbox{HERA-B} operation \cite{agingref}. Compared to Ar-CO$_{2}$
70/30 it also has larger primary ionization, 24 primary electrons in average
for a minimum ionizing particle in the 3 mm gas gap for Ar-DME compared to 18 for Ar-CO$_{2}$ which has
also significantly larger transverse diffusion. However, there are also disadvantages 
of Ar-DME:
The Ar-DME mixture is flammable and constitutes therefore a safety risk. DME is
not commercially available with guaranteed purity. Thus every bottle would have to be checked carefully in
aging tests to ensure that the detectors are not polluted. In addition,
we realized that DME is absorbed by Kapton leading to a reduction of the GEM
tension with time. After flushing the chamber for 2 weeks, we observed a sag of a foil by 800 $\mu$m.
The original tension could be recovered by changing the gas \cite{hild99}. The use of DME would have required additional
constructional efforts to keep the GEM foil in place.

\begin{figure}[tbp]
\begin{center}
\epsfig{clip=,file=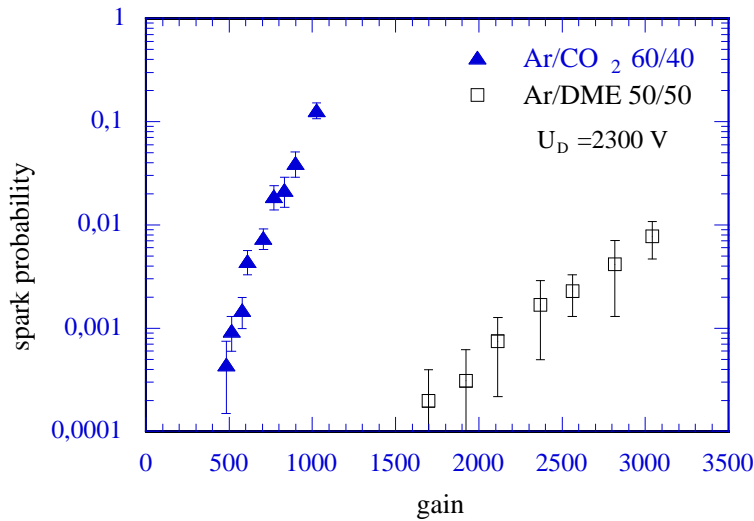,width=10cm}
\end{center}
\caption{{Discharge rate induced by $\protect\alpha$-particles in a
MSGC versus gas gain (arbitrary units) for two gas mixtures Ar-DME and Ar-CO$_{2}$.}}
\label{co2sparks}
\end{figure}

Comparative sparking tests in the laboratory with $\alpha $ particles
in the gas indeed demonstrated that the spark protection of
Ar-CO$_{2}$  is significantly worse for the same gas gain. This
is illustrated in Figure \ref {co2sparks} which shows the spark rates
for Ar-CO$_{2}$  and Ar-DME  versus gas gain. The result cannot be safely extrapolated to the
low operation voltages in a MSGC/GEM detector. It is nevertheless
obvious that Ar-CO$_{2}$ offers less safety with respect to gas
discharges than Ar-DME. Protection against sparking can be strongly
improved for Ar-CO$_{2}$ by adding small amounts of water to the gas.
However, as discussed below, the addition of water led to fast gas aging  and was therefore ruled
out for \mbox{HERA-B}.

\subsection{Experimental method for gas aging tests}

Big efforts have been made by several groups to study gas aging phenomena in
MSGCs \cite{rd28}. These studies are difficult and hard to reproduce
because gas aging depends on a multitude of different parameters. It was
however clear from the beginning that the MSGC structures are very
vulnerable and that even tiny amounts of pollutions by organic materials can
cause fatal deposits on the anodes. Therefore, most of the counting gases
which have been used in wire chambers involving organic quenchers cannot be
used in MSGCs for high rate applications. Moreover, we made a very careful
selection of materials like composites, glues and applied elaborated
cleaning and outgassing procedures to avoid pollution of the counting gas.
The gas system was entirely based on stainless steel tubing.
The sealing rings were made of Kalrez\footnote{Dupont Dow Elastomers, Belgium.}, 
materials like Teflon were avoided.  All materials were also tested for
out-gassing both using gas chromatography and by exposing large surfaces of them to the
input stream of counting gas used in gas aging tests of MSGC detectors.

An X-ray tube with copper anode was used to produce a high flux of photons.
The X-rays were collimated by a lead collimator such that a round area of 113 mm$%
^{2}$ was irradiated. The detector was operated at a total gas gain of about 3500. The gas
flow was arranged such that the gas in the detector was exchanged twice per
hour. The typical counting rate was about 100 kHz/cm strip length. One X-ray
count led to an average avalanche size of about 840000 $e^{-}$ and this rate
corresponds to a charge deposition rate per cm strip length which is about
20 times higher than that of the hottest region at \mbox{HERA-B}. This acceleration
factor was varied between 1 and 40 for different tests but the large
integrated charge depositions with illuminations over several months used
acceleration factors of not more than 20. The position of the Cu $%
K_{\alpha }$ line was evaluated in regular intervals during these measurements
to track possible changes in the gas gain.

 Aging tests were carried out with a large number of detectors over
a period of three years. With the full size \mbox{HERA-B} MSGC/GEM pre-series detectors we 
observed neither for Ar-DME nor for Ar-CO$_{2}$ any significant change of 
the pulse height  up to an integrated charge of
45 mC/cm which corresponds to about 6 years of \mbox{HERA-B} operation.
This result was supported by optical inspection
of the irradiated areas which showed no sign of depositions on the anodes.
This result was interpreted as sufficient reassurance for our choice of
counting gas and constructional materials to start mass production of
the detectors in spring 1998. It should be noted that our result was in line with similar
measurements of other groups \cite{agingref}. All their results were
 obtained with small irradiated areas of sometimes only
7 mm$^{2}$.

\subsection{ Observation of gas aging for Ar-DME mixtures}

\begin{figure}[tbp]
\begin{center}
\epsfig{clip=,file=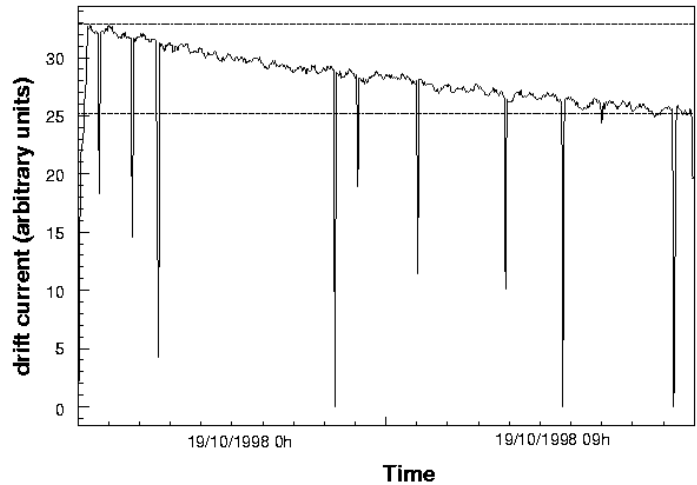,angle=-90,width=10cm}
\end{center}
\caption{Drift current (arbitrary units) as measured in the MSGC/GEM detector versus
operation time during intense proton irradiation at PSI. The beam current
was kept stable during the exposure.The gaps correspond to beam losses.}
\label{psicurrent}
\end{figure}

\begin{figure}[tbp]
\begin{center}
\epsfig{clip=,file=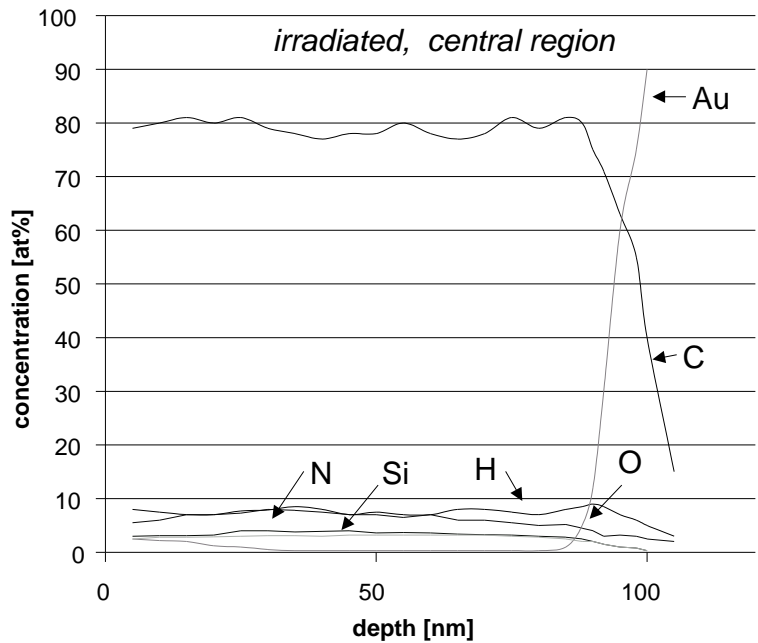,width=10cm}
\end{center}
\caption{{Concentration of elements as a function of depth for the
anode area in the irradiated region. A layer of carbon of 80 nm thickness is
covering the gold surface.}}
\label{fraunhoferanalyse}
\end{figure}

\begin{figure}[tbp]
\begin{center}
\epsfig{clip=,file=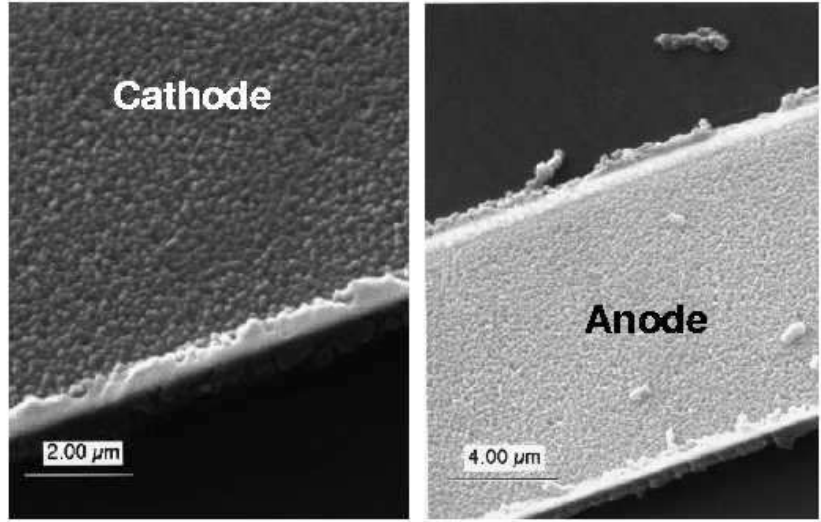,width=12cm}
\end{center}
\caption{{Photograph (electron microscope) of anodes in the irradiated
area of the MSGC after beam test at PSI.}}
\label{psideposit}
\end{figure}

In autumn 1998 a beam test was carried out at PSI with two full size MSGC/GEM
detectors using Ar-DME 50/50 as counting gas. The goal was twofold: during
daytime the pion beam was used to study efficiencies, resolution and the
performance of our trigger chain. During night time the detectors were
exposed to an intense proton beam to accumulate a higher charge dose. The
beam spot had a Gaussian profile with a typical full width at half maximum of \mbox{5
cm}. Within few days a total accumulated charge of \mbox{2 mC/cm} was reached in the
center region. We observed a rather fast reduction of the drift current
during proton irradiation for constant beam current. This was even seen
online as shown in Figure \ref{psicurrent} for a 12 hour period of proton
running at maximal intensity (about 20 times \mbox{HERA-B} charge deposition rate).
The detectors were subsequently investigated in the  laboratory.
Within the illuminated area the detectors still showed
acceptable gas gain but outside the illuminated area they did not count at all.
Optical inspection showed severe deposits on those anodes
which were connected to ground and therefore experienced the nominal electrical field.
Disconnected anodes showed no deposits. Depositions were found
both in the irradiated area and over the rest of the detector. Parts
of an irradiated wafer were sent to the Fraunhofer Institut f\"{u}r Schicht-
und Oberfl\"{a}chentechnik which carried out a surface analysis shown in
Figure \ref{fraunhoferanalyse}. The anodes in the irradiated area were
covered by a layer of \mbox{80 nm} pure carbon. A picture of
these deposits as seen with an electron microscope is shown in Figure \ref
{psideposit}. The areas outside, which were not
irradiated were covered by an insulating layer of about \mbox{35 nm} composed of
hydrogen and carbon. 

\begin{figure}[tbp]
\begin{center}
\epsfig{clip=,file=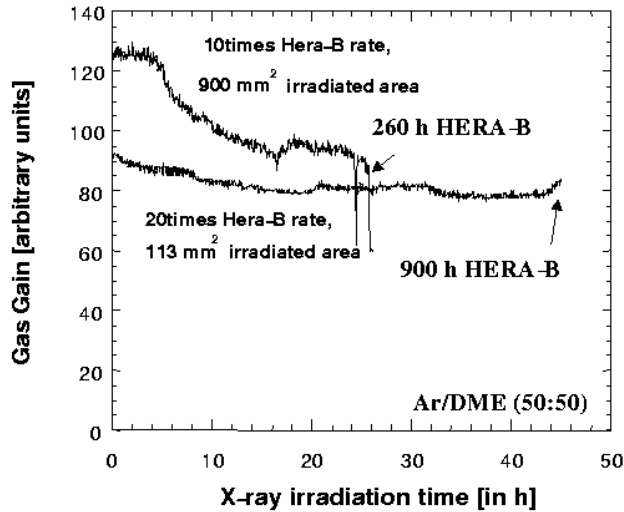,width=9cm}
\end{center}
\caption{{Gas gain versus irradiation time for a MSGC/GEM detector
using Ar-DME 50/50 and two different sizes of the irradiated area. The
detector was irradiated with X-rays using a charge deposition rate
corresponding to 10 respectively 20 times the maximal rate expected at \mbox{HERA-B}.}}
\label{dmeaging}
\end{figure}

This unexpected result prompted a series of new laboratory aging tests. The
final outcome is very surprising. The Ar-DME gas mixture shows gas aging
under X-ray irradiation if the area which is irradiated is large enough. 
This fact is illustrated in Figure \ref{dmeaging} which shows the gas
gain versus irradiation time for a collimator with an area of 113 mm$^{2}$ as
used before compared to one with an area of \mbox{900 mm$^{2}$}. The large 
area was irradiated at half the local intensity of the smaller one.
Identical chambers operated with Ar-CO$_{2}$ show no aging independent 
of the size of the irradiation area.  

Apparently, irradiation of a
large area with Ar-DME as a counting gas leads to fast aging and visible deposits
on the anodes. These deposits are not limited to the irradiated area. 
We therefore have to draw the conclusion that gas aging effects
depend strongly on the size of the irradiated area. All evidence supports the
assumption that the origin of these deposits is the DME itself. 
We have no explanation for this experimental
fact but it invalidates all aging tests done before.

Subsequently Ar-CO$_{2}$ was chosen as the main candidate for the counting gas.

As this gas mixture has low primary ionization and large transverse electron diffusion
leading to large cluster width in the detector, we tested several other 
gas mixtures without hydrocarbons.  These were Ne-CO$_{2}$, Kr-CO$%
_{2}$ and He-CO$_{2}$ mixtures. The results were unfortunately discouraging.
Measurements with MSGCs showed that for the same avalanche charge on the
anodes these gases led to comparable cluster widths but poorer
performance with respect to discharge protection. We therefore chose the
Ar-CO$_{2}$ 70/30 gas mixture.

\subsection{Aging results for different materials and counting gases}

As explained in Section 3, the safety against gas discharges became a
major concern. We therefore tested different combinations of electrode materials and counting gases
which promised better protection.

\begin{figure}[tbp]
\begin{center}
\epsfig{clip=,file=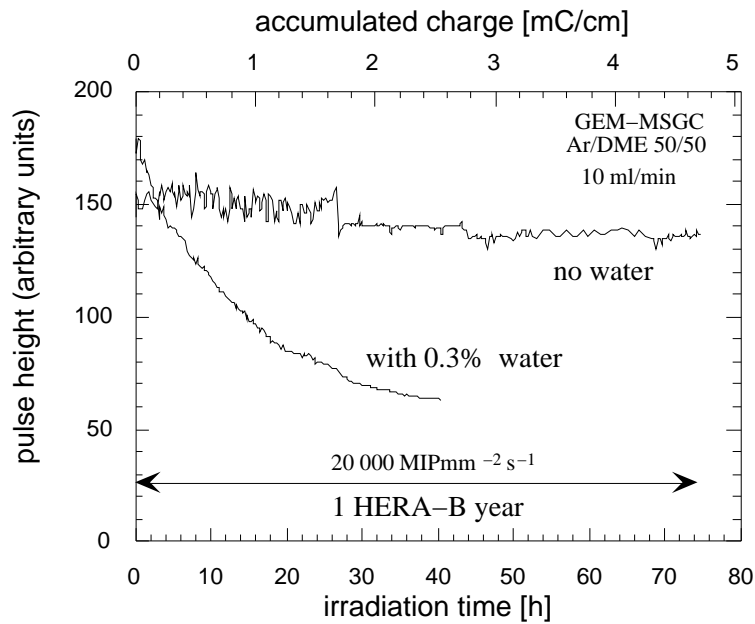,width=10cm}
\end{center}
\caption{Development of the pulse height for a gas mixture of
Ar-DME with 0.3\% of water and without water as function of the
irradiation time.}
\label{wateraging}
\end{figure}

The operation characteristics of a MSGC can be significantly improved by
adding small amounts of water, typically 0.3\%, to the counting gas. This
avoids discharges near insulators e.g. near the detector frames and anode
ends and improves the overall protection against sparking, probably by reducing 
the surface resistivity of plastic materials which tend to absorb water.
Addition of water to the counting gas shows however very severe and fast gas
aging for both, Ar-DME and Ar-CO$_{2}$. This is shown in Figure \ref
{wateraging} which compares the pulse height as a function of accumulated
charge for Ar-DME with and without an admixture of water. For the gas
with water admixture the pulse height drops by a factor three for an accumulated
charge of less than \mbox{3 mC/cm}. Moreover the anodes show deposits in the
irradiated area.

We also tested MSGCs with aluminium electrodes. Aluminium electrodes are easier to
manufacture and sparking tests showed that the electrodes are more robust
against discharges than gold but aluminium is known to introduce aging in MSGCs \cite{bateman}. 
Our aging tests with aluminium MSGC/GEM
detectors confirmed that this material cannot be used 
with Ar-DME or with Ar-CO$_{2}$. In both cases we observed a fast reduction
of the pulse height by more than 20\% for an accumulated charge of only \mbox{2.7
mC/cm}. Moreover, the electrodes were severely damaged. Especially the
cathodes showed bubbles and craters in the irradiated area. 

\begin{figure}[tbp]
\begin{center}
\epsfig{clip=,file=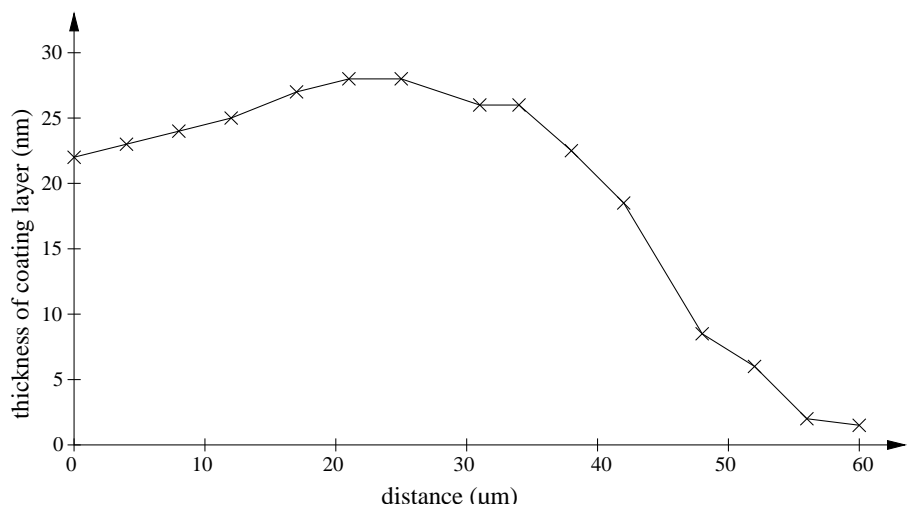,width=12cm}
\end{center}
\caption{{Thickness of the coating layer across the cathode anode gap after strong irradiation.
 The anode starts at 60 $\mu$m.}}
\label{diamondaging}
\end{figure}

A rather interesting phenomenon was observed for very long irradiation times for chambers operated
with gas mixtures without indication for gas aging like Ar-CO$_{2}$ and Ar-CO$_{2}$%
-CF$_{4}$ which is used in the Outer Tracker of HERA-B. After a stable period 
corresponding to about one \mbox{HERA-B} year
these chambers show a strong increase of pulse height with time of up to a
factor four. For some intermediate time the gas gain becomes very
inhomogeneous, changing locally by more than a factor two. After an exposure
corresponding to about five years of \mbox{HERA-B} operation the gain becomes stable
again but is about four times higher than at the beginning. The gain in this
state depends strongly on the counting rate which indicates that the surface
near the anodes has become insulating such that charging-up affects the
gain. The origin of these effects was clarified by a detailed surface
analysis. The intense plasma which is created during the
aging tests destroys the DLC layer near the anodes. It is
gradually etched away up to the point that there is an insulating strip near
the anode edges. This is illustrated in Figure \ref{diamondaging} which
shows the measured thickness of the coating as a function of the
distance between anode and cathode for an area where a charge
of \mbox{40 mC/cm} was accumulated. The coating is etched away almost completely near
the anode and is reduced over the whole gap compared to the starting
thickness of about 80 nm. With Ar-CO$_{2}$ this effect is relatively slow
under the conditions of our aging tests, for a gas mixture including CF$_{4}$
the effect is much faster such that the coating would be destroyed
completely for a collected charge of only \mbox{2 mC/cm}. This gas can therefore not be
used. The observed etching effect is of course likely to depend
strongly on the local current and plasma density and could be very different
for the running conditions of the \mbox{HERA-B} experiment. It is therefore
impossible to predict if and how fast such a change would appear at \mbox{HERA-B}.
The final beam test at PSI in 1999 using low energy protons was used to
accumulate a deposited charge equivalent to 1/3 year of \mbox{HERA-B} operation.
During this relatively small exposure no signs of gas aging or damage of the DLC 
were observed.

\section{ Operation at \mbox{HERA-B}}

\subsection{Treatment of the detector components before installation}

During production, the different parts of the ITR went through
several tests:

The wafers were checked at Zurich directly after production (see Section 2.2). Interrupted
anodes were recorded. 

All GEMs were tested in Heidelberg before chamber construction.
The GEMs had to sustain 500 V in a nitrogen atmosphere for 12 hours.
        
The individual chambers were transported to Siegen for bonding of the electronics. 
At Siegen they were flushed with the final chamber gas and put 
to high voltage. Within a period of 24 hours the voltages were ramped up to 520 V 
for the cathode and to 410 V across the GEM. All currents were recorded and put 
into a reference file.
About half of the chambers showed anode cathode shorts. These shorts were
eliminated by disconnecting the anode. In 10 chambers fatal GEM shorts were
observed.

Chambers fulfilling the requirements went back to Heidelberg. There the stations were put 
together and fully assembled  stations were shipped to DESY.
Before installation another high voltage test similar to the one at Siegen
was performed. The required voltages were 490 V for the cathodes
and 400 V across the GEMs. In 12 out of 144 tested chambers an anode cathode
short was detected. One chamber showed a GEM short.

Apparently, each construction or transportation step introduced some
damage to part of the detectors and the intermediate high voltage tests
could not exclude additional problems occurring later in the hadron beam of
HERA-B.

\subsection{Operation conditions and observed problems}

During the period from July 1999 to May 2000 a total of 150 chambers were
installed and operated at \mbox{HERA-B}. The first few months of the year 2000 were
mainly devoted to commissioning of the detectors. Data were routinely taken
from April to the end of the running period by August 2000. 
The maximum operation time of 1100 hours was reached by 56 chambers. 
During the last two months 90 \% of
the installed chambers were routinely participating in data taking without
major problems.

\begin{figure}[tbp]
\begin{center}
\epsfig{clip=,file=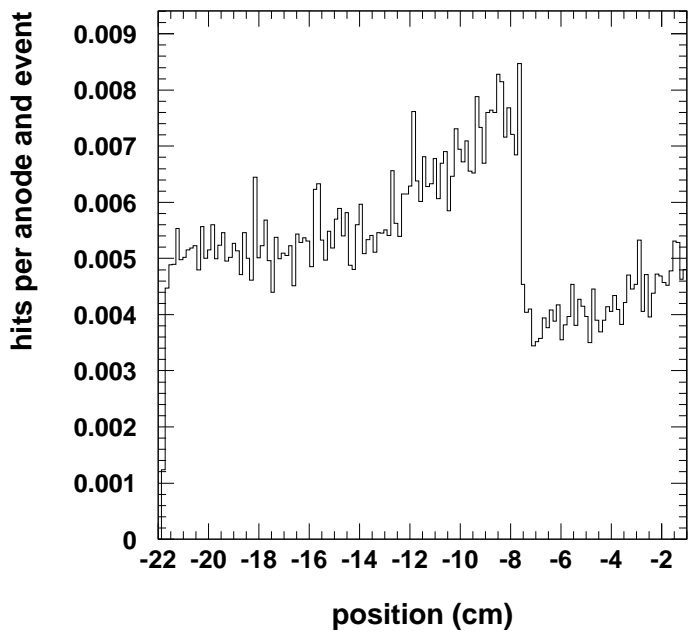,width=9cm}
\end{center}
\caption{Hit density as a function of the horizontal distance to the beam axis.}
\label{hitdist}
\end{figure}

The interaction rate at the target varied between 3 MHz and 40 MHz. Most of the data were taken
at 5 MHz. A typical distribution of the hit density in a detector is shown in Figure \ref{hitdist} for 5 MHz interaction rate.
The hit density is about 0.006 per
anode and event. The shape of the distribution is determined by the beam
recess of the MSGC wafer (see Fig. 1) and the radial decrease of particle density.

The raw pulse height distribution cannot be used to determine the gas gain because it is
dominated by soft signals which do not belong to tracks. We therefore need to 
reconstruct tracks first to measure a signal over noise distribution and the efficiency.
Since at the beginning of the run the track reconstruction program was not yet well tuned, 
it was possible only for the
chambers in front of the magnet to estimate during running the efficiencies which have to be known in order 
to adjust the high voltage. Typical voltage settings were -510 V for the cathode and  
420 to 450 V for the GEM.
Individual GEM voltage settings were necessary to correct for the
gain variations of up to a factor 2.5 from GEM to GEM which are caused by
slight alignment imperfections of the copper holes at the two sides of the
GEM. The high
voltage settings for the remaining chambers could not be adjusted and 
therefore the chambers were operated at GEM voltages below 440 V and 
correspondingly more moderate gain. 

While at the end of the data taking period running was rather smooth,
several problems occurred after installation:

\begin{itemize}
\item  Four GEMs developed conductive paths between their two copper surfaces. The resistivity 
dropped to values of the order of M$\Omega$, values which effectively short the two GEM electrodes.

\item  In 43 \% of the chambers the MSGC electrodes showed one or two short circuits 
between anode and cathode strips.
\end{itemize}

These problems occurred shortly after installation and were probably 
due to spurious dust or production defects.

\subsubsection{GEM sparks}

\begin{figure}[tbp]
\begin{center}
\epsfig{clip=,file=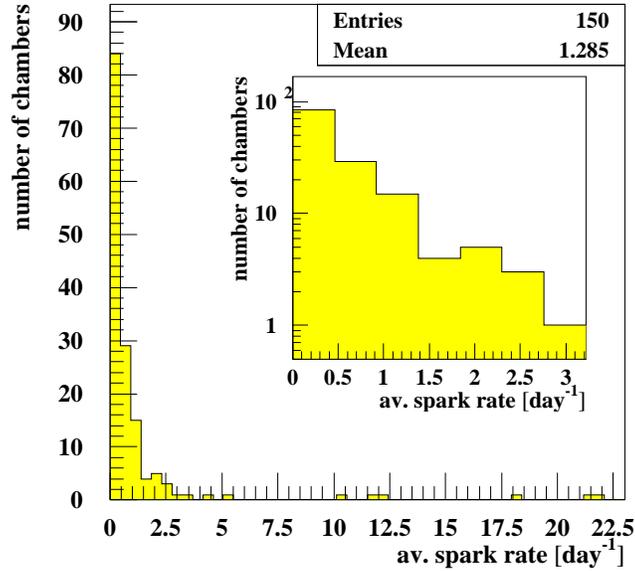,width=9cm}
\end{center}
\caption{{Number of chambers as a function of their average spark rate. 
Most chamber show rates below two sparks per day at an average target rate of 5 MHz.}}
\label{meanspark}
\end{figure}

Sparking in GEMs cannot be avoided completely, but under normal conditions
the rate is quite low, of the order of a single spark per day (see \mbox{Figure 
\ref{meanspark}}), a rate which is tolerable. Only six chambers showed
sparking at rates above ten per day.

\begin{figure}[tbp]
\begin{center}
\epsfig{clip=,file=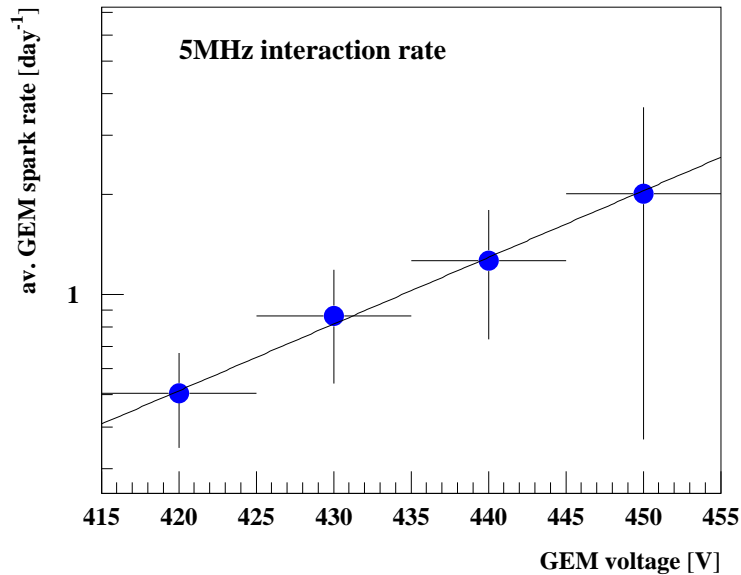,width=10cm}
\end{center}
\caption{{Spark rate versus GEM voltage averaged over all chambers with
rates below six per day. The error bars include systematic uncertainties.}}
\label{sparkvsgem}
\end{figure}

The spark rate increases exponentially with the GEM voltage rising from 1
to 5 sparks/day when the GEM voltage is increased from 400 V to 450 V. This
is shown in Figure \ref{sparkvsgem} for chambers behaving normally. Chambers
with exceptional high rates are excluded in this Figure to avoid a biased
picture. As expected, the sparking rates increased with the beam
intensity. Similar results have been reported in Refs. \cite{zieg01,bach00} where
a linear dependence of the sparking probability on the beam intensity and a 
strong dependence on the beam composition has been
observed.

While under normal conditions the spurious sparking does not disturb the
performance of the detectors, in some cases continuous sparking produced GEM
``shorts'' by carbonizing the polyimide surface in the hole.

In some cases when the resulting conductivity across the GEM was low, the
``short'' could be burned away by applying a short current pulse which
evaporates the copper near the affected hole. When this was not possible, the
whole chamber was lost.

At one occasion, the damage of a GEM could be correlated to a spike in the
target rate.

\subsubsection{Anode cathode shorts}

A total of 79 anode cathode shorts were observed during the run period in 2000. Due to the HV grouping of
the cathodes this corresponds to a loss of 1.1\% of all anode strips. Out of
the total, 29 shorts were produced already within the first 10 hours
after switching on and only 16 occurred in the second half of the running
period. Most of the late shorts coincided with an increase in the high
voltage.

Inspection of the chambers during the HERA luminosity upgrade shut-down revealed that some
shorts are correlated to anodes with lithographic defects. Interrupted
anodes show a tenfold higher probability to produce a short than perfect
anodes. Interrupted anodes have now been disconnected electrically from ground to
avoid this problem. The corresponding loss in efficiency is
completely negligible.

\subsubsection{Consequences}

Most problems occurred at the beginning of the operation in the beam. They
can be avoided or at least reduced to a tolerable level by the following
measures:

\begin{itemize}
\item  operating the detectors for some days at overvoltage before
installation,

\item  carefully training the chambers in the \mbox{HERA-B} beam before applying
the full high voltage,

\item  HV control based on monitoring of the currents and on spark detection,

\item applying a coordinated ramping of the different chamber high voltages,
 
\item  avoiding unstable beam conditions.
\end{itemize}

During the 2000 running period a training procedure has been worked out
which has led to a considerable improvement of the reliability of the
detectors. The voltages are raised slowly in ten steps over a period of two
to four weeks (400 hours beam time) depending on the chamber behavior. Very few
problems were observed with chambers treated in this way.

A control system based on microprocessors inside the high voltage distribution
boxes has been optimized. It continuously monitors the drift and
cathode currents and the GEM voltages. In addition, it detects sparking through rapid voltage changes. When a
spark occurs, the voltages are decreased thereby avoiding sequential
sparks. After some delay, the voltages are automatically ramped up
again. The behavior of all chambers is continuously monitored online.

Meanwhile all damaged chambers have been exchanged. Cathode groups have been
re-activated by cutting the anode responsible for the short. Extrapolation
from the experience gained last year indicates that we should be able to
operate the ITR with less than 1\% of channels lost per year.

\subsection{Performance}

\begin{figure}[tbp]
\begin{center}
\epsfig{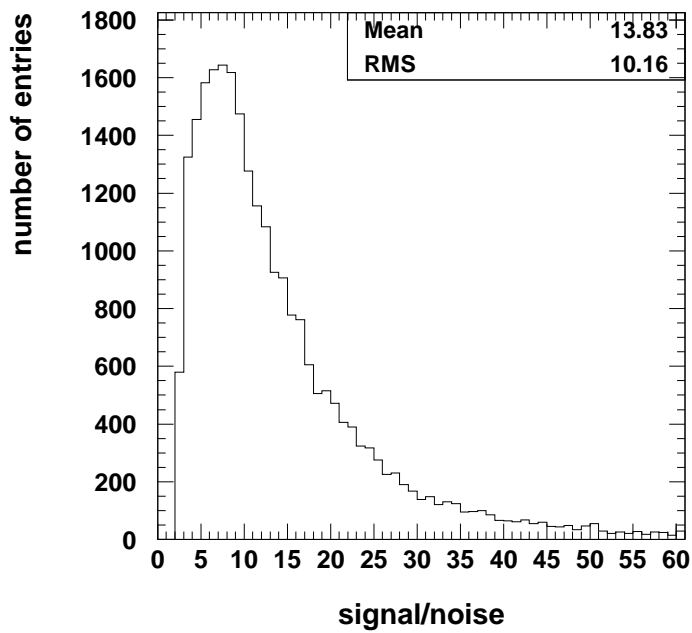}
\end{center}
\caption{{Signal over noise distribution.}}
\label{s_ndist}
\end{figure}

The detectors show a noise distribution with a mean value of 2500 electrons. 
This value is well
compatible with the estimate from the strip capacitance and the amplifier input
resistor value. A signal over noise (S/N) plot is shown in Figure 
\ref{s_ndist}.

\begin{figure}[tbp]
\begin{center}
\epsfig{clip=,file=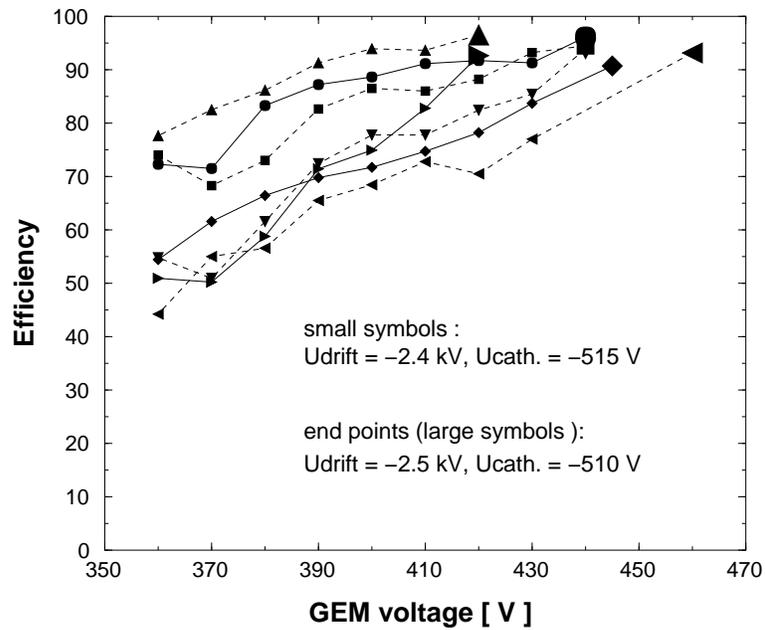,width=10cm}
\end{center}
\caption{{Efficiency of seven detectors of station MS01 versus GEM voltage
for fixed cathode and drift voltages. The large symbols show the
efficiencies for the current voltage settings after individual gain
adjustments.}}
\label{effhera}
\end{figure}

Efficiencies of the chambers in front of the magnet could be determined
during running using well reconstructed tracks extrapolated  from the vertex detector.
They are presented in Figure \ref{effhera} as a function of the GEM voltage.
The efficiencies are above 90 \% which is sufficient for tracking.

As mentioned above, there are considerable variations in the GEM performance
from chamber to chamber which are compensated by adjusting the GEM high voltage 
which can be individually set for each chamber. The local gain
variations across a single chamber, however, are quite small, namely below
10 \%. 


\begin{figure}[tbp]
\begin{center}
\epsfig{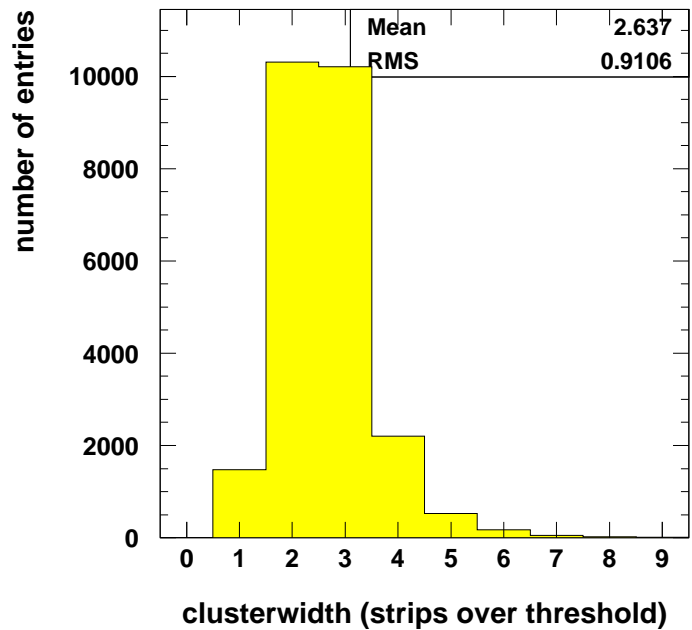}
\end{center}
\caption{{Distribution of cluster width.}}
\label{stripmult}
\end{figure}

The strip multiplicity distribution shown in Figure \ref{stripmult}
corresponds to a mean value of 2.6 for a threshold of 5000 electrons, 
which is twice the average noise level. This rather high
value is compatible with the large transverse diffusion in Ar-CO$_{2}$ along the
long drift path of the electrons.

\begin{figure}[tbp]
\begin{center}
\epsfig{clip=,file=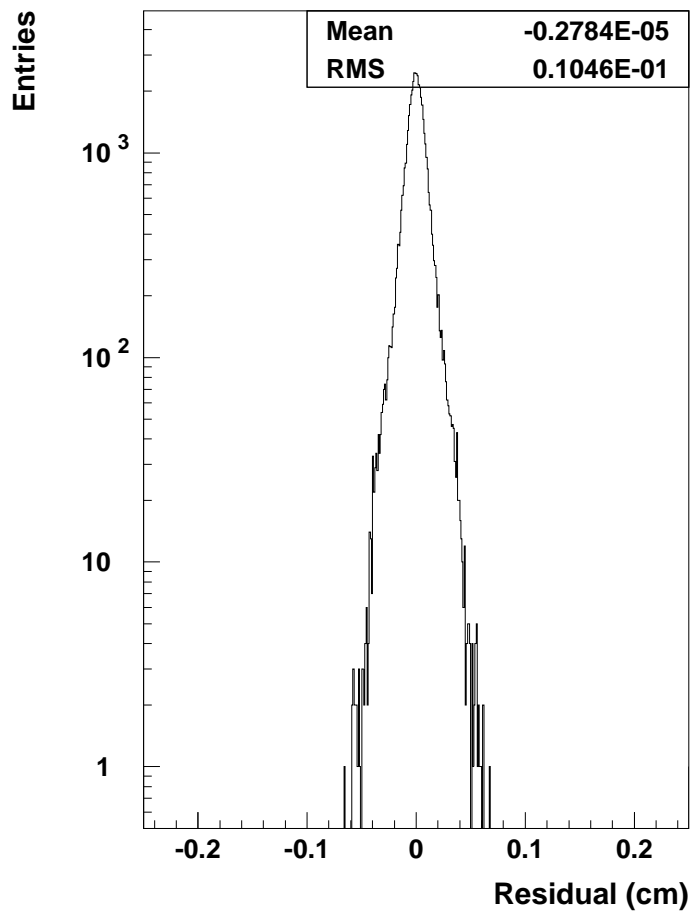,width=9cm}
\end{center}
\caption{{Distribution of track residuals.}}
\label{residuals}
\end{figure}

The ITR chambers have been geometrically aligned using an iterative fitting method. 
A residual distribution for fitted tracks of particles with momenta between 35 and 50 GeV/c is presented in
Figure \ref{residuals}. In average 15 chambers contribute to one track. From a comparison 
of the observed root mean error of \mbox{105 $\mu$m} with a Monte Carlo simulation we conclude 
that the intrinsic spatial resolution of the detectors is better than \mbox{110 $\mu$m}. 

\section{Conclusions}

After many unexpected difficulties the development of the ITR for \mbox{HERA-B} has led 
to a detector which is working to the required specifications.
Efficiencies greater than 90\% were achieved for part of the
chambers and are expected for the full detector during the next running
period. Extrapolation of the present experience indicates that
inefficiencies due to anode cathode shorts will remain below the 1\% level.

Irradiation with X-rays have revealed that aging for Ar-DME gas mixtures
depend strongly on the size of the irradiated area. 
Measurements with Ar-CO$_{2}$, 70/30 did not show the same
behavior and indicate that our detectors would adequately perform for more
than six \mbox{HERA-B} years. Irradiation also affects the DLC
which is etched away starting from the anode borders. This leads to gain
variations with time and beam intensity.

The MSGC with the dimensions necessary at \mbox{HERA-B} and the beam conditions of
HERA cannot be operated reliably without the addition of a GEM. At the gas
amplification required to obtain sufficient efficiency the spark rates would be
lethal. The situation here is different from that at LHC where with much smaller
detectors simpler solutions would have been possible \cite{cmsmilestone}. Our
system is quite fragile and expensive and requires a complicated HV
steering. With the knowledge gained in the last few years, it has become
clear that new, more robust devices like multi GEM detectors \cite{multigem,zieg01,Ketzer}
are better suited for harsh beam environments than MSGC based systems.

\section*{Acknowledgment}
We acknowledge the strong effort
of the technical stuff of the collaborating institutions. In particular, 
we wish to thank S. Henneberger, R. Rusniak, A. Rausch, Ch. Rummel, D.
Gieser, R. Eitel and S. Rabenecker (Heidelberg), O. Meyer, G. Schmidt (Siegen), 
K. Boesiger, K. Esslinger, B. Schmid and S. Steiner (Zuerich), Serguei Cheviakhov (DESY). Our
work profited very much from the close collaboration with the vertex
detector group of the MPI Heidelberg, especially in the areas of readout and monitoring.

\end{document}